\documentclass{aa}
\usepackage{graphicx}
\usepackage{txfonts}
\usepackage{url}

\usepackage{amsmath}
\usepackage{subfigure}
\usepackage{graphicx}
\usepackage{enumitem}

\begin{document}

\title{123--321 Models of Classical Novae} 

\author{Jordi Jos\'e \inst{1,2}
   \and Steven N. Shore \inst{3}
   \and Jordi Casanova \inst{2} }

\offprints{J. Jos\'e}

 \institute{Departament de F\'\i sica, EEBE,
            Universitat Polit\`ecnica de Catalunya, 
            c/Eduard Maristany 10, 
            E-08930 Barcelona, 
            Spain\
            \email{jordi.jose@upc.edu}
            \and 
            Institut d'Estudis Espacials de Catalunya, 
            c/Gran Capit\`a 2-4, 
            Ed. Nexus-201, E-08034 Barcelona, 
            Spain\
            \and
            Dipartimento di Fisica ``Enrico Fermi'',
            Universit\`a di Pisa and INFN, Sezione di Pisa, 
            Largo B. Pontecorvo 3, I-56127 Pisa, Italy
}
       
\date{\today}

\abstract{High-resolution spectroscopy has revealed
large concentrations of  CNO and sometimes other intermediate-mass elements (e.g., Ne, Na, Mg, or Al, for ONe novae) in the shells ejected during nova outbursts,
suggesting that the solar composition material transferred from the secondary mixes with the outermost layers of the underlying white dwarf during the thermonuclear runaway.
  Multidimensional simulations 
have shown that Kelvin-Helmholtz instabilities provide self-enrichment of the accreted envelope 
with material from the outermost layers of the white dwarf, at levels that agree with observations. However, 
the Eulerian and time-explicit nature of most multidimensional codes used to date and the overwhelming computational load 
have limited their applicability,  and no multidimensional simulation has been conducted for a full nova cycle.
This paper explores a new methodology that combines 1--D and 3--D simulations. 
The early stages of the explosion (i.e., mass-accretion and initiation of the runaway) have been computed 
with the 1--D hydrodynamic code {\tt SHIVA}.  When convection extends throughout the entire envelope, 
the structures for each model were mapped into 3--D Cartesian grids and were subsequently followed with the multidimensional code {\tt FLASH}. 
Two key physical quantities were extracted from the 3--D simulations and subsequently implemented into {\tt SHIVA},  
which was used to complete the simulation through the late expansion and ejection stages:  the time-dependent amount of mass dredged-up from
the outer white dwarf layers, and the time-dependent convective velocity profile throughout the envelope.
This work explores for the first time the effect of the inverse energy cascade that characterizes 
turbulent convection in nova outbursts.  More massive envelopes than those reported from previous models with pre-enrichment have been found.
This results in  more violent outbursts, characterized by higher peak temperatures  and greater ejected masses, with metallicity enhancements in agreement with observations.}


\maketitle

\section{Introduction}
\label{sec:intro}
Multiple complementary approaches undertaken in the study of the nova phenomenon 
(i.e., spectroscopic determinations of chemical abundances, photometric studies of light curves, and  
hydrodynamic simulations of the accretion, explosion and ejection stages) have paved the way for our current 
understanding of these cataclysmic events (see Starrfield et al. 2008, 2012, 2016, Jos\'e \& Shore
2008, Jos\'e 2016, for reviews).  The canonical scenario  assumes a white
dwarf star as the site of the explosion in a short period, stellar binary system
 (with orbital periods mostly ranging between 1.5 hr and 15 hr; Diaz \& Bruch 1997). 
 The low-mass stellar companion (frequently a K-M main sequence star although 
observations demonstrate the presence of more evolved companions) overfills its Roche lobe, and 
matter flows through the inner Lagrangian point of the system.
The matter transferred (typically at a rate in the range $\dot M \sim 10^{-8} - 10^{-10}$ M$_\odot$ yr$^{-1}$) 
does not fall directly onto the compact star. Instead, it forms an 
accretion disk that orbits around the white dwarf. A fraction of this hydrogen-rich disk  
drifts inward and ultimately ends up on top of the white dwarf, where it is gradually compressed in semi-degenerate 
conditions to high densities. Compressional heating increases the temperature 
at the envelope's base until nuclear reactions set in and a thermonuclear runaway ensues. As a result, about  
$\sim 10^{-7} - 10^{-4}$ M$_\odot$ of nuclear-processed material is expelled into the interstellar medium 
at typical velocities of several $10^3$ km s$^{-1}$.
Such nova outbursts are quite common, constituting the second, most frequent type of stellar thermonuclear
explosion in our Galaxy, after type I X-ray bursts. Although only a handful, 5 to 10, are discovered per year, a much higher 
rate, around $50^{+31}_{-23}$ yr$^{-1}$, has been predicted from extrapolation of Galactic and 
extragalactic data\footnote{About 400 novae have been discovered in the Milky Way.  A catalogue of all 
novae detected in M31 (1159 novae), M32 (5), M33 (53), M81 (231), NGC 205 (4), and the Magellanic clounds (77), 
can be found at \url{http://www.mpe.mpg.de/~m31novae/opt/index.php?lang=en}.} (Shafter 2017).
%can be found at {\tt http://www.mpe.mpg.de/$\texttildelow$m31novae/opt/index.php?lang=en}.} (Shafter 2017).

The thermonuclear nature of nova explosions was first hypothesized by Schatzman (1949, 1951). This was followed by 
a number of significant contributions in the 1950s and 1960s (Cameron 1959, Gurevitch \& Lebedinsky 1957), 
including pioneering attempts to mimic the explosion through the coupling of radiative transfer in an optically 
thick expanding shell with hydrodynamics (Giannone \& Weigert 1967, Rose 1968, Sparks 1969, Starrfield 1971a,b).
Most of the modeling efforts to date have relied on one--dimensional (1--D) or spherically symmetric, hydrodynamic codes (see, e.g., 
Starrfield et al. 1972, 2016, Prialnik et al. 1978, Yaron et al. 2005, Hillman et al. 2016, Jos\'e \& Hernanz 1998, 
Jos\'e 2016, Denissenkov et al. 2013, Rukeya et al. 2017, and references therein). While this  approach can 
qualitatively reproduce most of the observational features of a nova outburst, it is increasingly clear
that the assumption of spherical symmetry cannot resolve a suite of critical issues, such as the way a
thermonuclear runaway (TNR) initiates (presumably as a point-like or multiple-point ignition) and propagates throughout 
the envelope (see Shara 1981, for pioneering work on localized, volcanic-like TNRs). Moreover, it has been realized that to reproduce the specific amount of mass ejected, 
the energetics of the event, and the chemical composition of the ejecta, a different approach was somehow required. 
In fact, while the material  accreted on the white dwarf is, in many cases, expected to be of solar composition (i.e., with a 
metallicity Z $\sim$ 0.02), the chemical abundance patterns spectroscopically inferred in the ejecta reveal
large amounts of intermediate-mass elements, resulting typically in Z $\sim$ 0.2 - 0.5. Mixing at the core-envelope 
interface has been regarded as the most likely explanation for such metallicity enhancements. 
Several mixing mechanisms have been proposed and explored 
to date in 1--D\footnote{Mixing by resonant gravity waves on the white dwarf surface has also been studied in 2--D 
(see Rosner et al. 2001, Alexakis et al. 2004). However, a very high shear must be imposed  to yield significant mixing, 
with a specific velocity profile.}, 
such as diffusion-induced mixing (Prialnik  \& Kovetz 1984; 
Kovetz \& Prialnik 1985; Fujimoto \& Iben 1992; Iben et al. 1991, 1992), shear mixing (Durisen 1977, 
Kippenhahn \& Thomas 1978, MacDonald 1983, Livio \& Truran 1987, Fujimoto 1988, Sparks \& Kutter 1987, 
Kutter \& Sparks 1987, 1989), and convective overshoot-induced flame propagation (Woosley 1986),
but none has succeeded in reproducing the range of metallicity enhancements inferred from observations 
(Livio \& Truran 1990). 

More promising results have been obtained by relaxing the constraints imposed by strict sphericity in the codes.
Multidimensional simulations of mixing at the core-envelope interface during 
nova outbursts 
have shown that Kelvin-Helmholtz instabilities can naturally lead to self-enrichment of the accreted envelope 
with material from the outermost layers of the white dwarf, at levels that agree with observations 
(Glasner \& Livne 1995, Glasner et al. 1997, 2007, 2012 Casanova et al. 2010, 2011a,b, 2016, 2018). In particular, 
3--D simulations (Casanova et al. 2011b, 2016) have provided hints on the nature of the highly fragmented, 
chemically enriched and inhomogeneous nova shells, observed in high-resolution. This, as predicted by turbulence 
theory (see Pope 2000, Lesieur et al. 2001), has been interpreted as a relic 
of the hydrodynamic instabilities that develop during the initial ejection stage.

The codes used for these multidimensional simulations (i.e., {\tt FLASH} and
{\tt VULCAN}) rely on {\it time-explicit} schemes. In general, partial differential equations involving 
time derivatives can be discretized in terms of variables determined 
at the previous time ({\it explicit} schemes) or at the current time ({\it implicit} schemes).
Explicit schemes are usually easier to implement than implicit schemes.
However, such schemes face severe constraints on the maximum time-step allowed, 
given by the Courant--Friedrichs--Levy (CFL) condition, to
prevent any disturbance traveling at the speed of sound from traversing more than one
numerical cell, which may lead to unphysical results (Richtmyer \& Morton 1994).
In contrast, implicit schemes allow longer time-steps, with no
preconditions, but they require an iterative procedure to solve the system of equations at each step.
The limitations posed by the CFL condition on the time-step 
make it difficult for explicit schemes to simulate the hydrostatic stages in the life of stars,
since an incredibly large number of time-steps would be required to this end. 
In this framework, all multidimensional simulations of mixing during
novae rely also on 1--D codes to compute the earlier, hydrostatic stages of the event
(i.e., mass-accretion). Only when the evolution of the star proceeds on a dynamical timescale is
the structure of the star mapped into a 3--D (or 2--D) computational domain 
 and followed by a suitable multidimensional code. The technique
has been named the {\it 1 to 3} (or $123$) approach.
To reduce the overwhelming computational load, simulations rely on
 small computational domains (i.e., a cube containing a fraction of the overall
star in 3--D; a box, in 2--D simulations). 
It is, however, worth noting that the {\tt VULCAN} code can naturally follow the expansion of the envelope, 
since it can operate in any combination of Eulerian and Lagrangian modes (see Glasner et al. 1997 and references therein, for details.). 
Frequently, a reduced nuclear reaction network is adopted, that
includes only a handful of species to approximately account for the energetics of the event. Parallelization techniques are implemented to distribute the computational
load among different processors (Martin, Longland \& Jos\'e 2018). 

All multidimensional simulations of mixing in novae reported to date rely on 
the {\it 123} technique to perfom {\it convection-in-a-box} studies, aimed at verifying
the feasibility of Kelvin-Helmholtz instabilities as an efficient mechanism of
self-enrichment of the accreted envelope with core material, for different chemical
substrates and white dwarf masses. The goal of this paper is to exploit the results
obtained in these multidimensional simulations (Casanova et al. 2016),
deriving prescriptions for the time-dependent convective velocity and mass dredge-up,
and inserting them back into a 1--D code to follow the final stages of the
outburst (i.e., {\it 3 to 1} [or $321$] approach, hereafter). 

The paper is organized as follows. The input physics and initial 
conditions of the simulations are described in Sect. 2. 
A full account of our 123-321 simulations for 
neon and non-neon (CO-rich) novae
are presented in Sect. 3. Finally, the significance of our  results and
our main conclusions are summarized in Sect. 4.

\section{Model and initial setup}
\label{sec:model}

\subsection{The 123 approach}

The early stages of the evolution of Models CO1 and ONe1 reported in this paper, the 
mass-accretion and initiation of the thermonuclear runaway, have been computed 
in spherical symmetry (1--D), with the time-implicit, Lagrangian, hydrodynamic code {\tt SHIVA}, extensively
used in the modeling of different stellar explosions (classical
novae, X-ray bursts, and subChandrasekhar supernovae). 
{\tt SHIVA} solves the standard set of differential
equations of stellar evolution in finite-difference form: conservation of mass,
momentum, and energy, and energy transport by radiation and
 convection. 
It relies on a time-dependent formalism for convective transport whenever 
the characteristic convective timescale becomes larger that the integration itime-step 
(Wood 1974).  Partial mixing between adjacent convective shells is treated by means of a
diffusion equation (Prialnik, Shara, \& Shaviv 1979).
The equation of state includes contributions
from the degenerate electron gas, the ion plasma, and
radiation. Coulomb corrections to the electron pressure are also taken
into account. Radiative and conductive opacities are considered in the
energy transport. Energy generation by nuclear reactions is obtained using
a network that contains
120 nuclear species, ranging from H to $^{48}$Ti, connected through 630 nuclear
interactions, with updated {\tt STARLIB} rates (Sallaska et al.
2013). See Jos\'e \& Hernanz (1998) and Jos\'e (2016) for further details 
on the {\tt SHIVA} code.

Spherical accretion of solar-composition material, 
at a constant rate of $2 \times 10^{-10}$ M$_\odot$ yr$^{-1}$, onto white dwarfs of 1 M$_\odot$ 
(assumed to be CO-rich) and 1.25 M$_\odot$ (ONe-rich), was assumed for Models CO1 and ONe1, respectively. 
When the temperature at the core-envelope interface
reached T$_{ce} = 10^8$ K, the structures for each model were mapped into 3--D Cartesian
grids and were subsequently followed with the multidimensional, parallelized, explicit, 
Eulerian code {\tt FLASH}. {\tt FLASH} relies on the piecewise parabolic interpolation of physical quantities
to solve the set of equations that characterize a stellar
plasma (Fryxell et al. 2000). The code uses adaptive mesh refinement to improve the accuracy in critical regions of the
computational domain. 

A 3--D computational domain of $800 \times 800 \times 800$ 
km$^3$, initially comprised of 112 unevenly spaced vertical (radial) layers and 512 equally spaced layers along each 
horizontal (transverse) axis, in hydrostatic equilibrium, was adopted for Model CO1 to conduct our {\it convection-in-a-box} simulations.
The masses of the regions mapped into the 3-D Cartesian grids followed with the {\tt FLASH} code are $6.75 \times 10^{-9}$ M$_\odot$ (envelope) and $1.80 \times 10^{-7}$ 
M$_\odot$ (outer white dwarf layers), for Model CO1, and  
$1.37 \times 10^{-8}$ M$_\odot$ (envelope) and 
$7.17 \times 10^{-7}$ M$_\odot$ (outer white dwarf layers), 
for Model ONe1. 
The maximum resolution adopted, with five levels of refinement, was 
$1.56 \times 1.56 \times 1.56$ km$^3$ to handle the sharp discontinuity at the core-envelope interface,
although a typical zoning of $3.125$ km was employed along each dimension during most of the simulation. 
For Model ONe1, a computational domain of $800 \times 800 \times 400$ km$^3$ was adopted, with  88 unevenly spaced vertical 
layers and the same equally spaced layers adopted for Model CO1 along each horizontal axis. The maximum and typical resolution adopted 
were the same as for Model CO1. In all models, 
periodic boundary conditions were implemented at lateral faces, while hydrostatic conditions were imposed through the 
vertical boundaries, reinforced with a reflecting
condition at the bottom and an outflow condition at the top (see Casanova et al. 2016, and references therein).

The choice of $T_{ce} = 10^8$ K as the condition for mapping the 1--D structure into a 3--D grid 
is based on the work reported by Glasner et al. (2007), who demonstrated the universality 
of mixing driven by Kelvin-Helmholtz instabilities, independent of the stage (i.e., time and temperature) at which the 1--D models are mapped  
(tests performed by Glasner et al. in 2--D included mapping at T$_{ce} =  5 \times 10^7$, 
$7 \times 10^7$, and $9 \times 10^7$ K).  Therefore, while mapping at earlier temperatures has no noticeable effect 
on mixing, the computational time 
increases dramatically because of the time-explicit nature of the {\tt FLASH} code.

In both CO1 and ONe1 models, an initial top-hat, 5\% temperature perturbation with a radius of 1 km is introduced, 
operating only during the first time-step near the envelope's base, to create fluctuations along the core-envelope 
interface. {\tt FLASH} divides each computational cell in a number of subcells and determines how many subcells are affected by the 1 km-perturbation. This results
in a single, fully perturbed cell ($1.56 \times 1.56 \times 1.56$ km$^3$), whose temperature is increased by 5\%, surrounded by a handful of neighboring shells with a smaller temperature
increase (the average temperature of each cell relies on the number of subcells affected by the perturbation). 
This induced strong buoyant fingering (see Casanova et al. 2011a, for details on the implementation and effect of the initial perturbation). The development of Kelvin-Helmholtz instabilities
drives an efficient dredge-up of outer core material into the envelope by the rapid formation of small convective 
eddies in the innermost envelope layers. Due to the Eulerian nature of {\tt FLASH}, calculations are stopped when 
the convective front hits the upper computational boundary. The simulations cannot be continued into most of the expansion and ejection stages.

\subsection{The 321 approach}

At the end of the 3--D simulations, two key physical quantities can be obtained: (i) the time-dependent amount of mass dredged-up from
the outer white dwarf layers into the solar accreted envelope, and (ii) the time-dependent convective velocity profile in the envelope.
These two quantities are essential ingredients for the {\it 321 approach}. 
Within mixing-length theory\footnote{Mixing-length theory implicitly imposes 
$\nabla \cdot (\rho v) \equiv 0$, since it considers only thermal transport and not mass fluxes; it is an equilbrium, mean field theory.}
 (Biermann 1932; see B\"ohm-Vitense 1958 and Cox \& Giuli 1968,
for early, detailed descriptions of the theory), the Schwarzschild criterion establishes that 
convective energy transport sets in whenever superadiabatic temperature
gradients occur, $\nabla > \nabla_{ad}$, where  $\nabla \equiv d\ln T/d\ln P$.
 
In each convective shell, the convective energy flux, $F_{conv}$, can be classically expressed as
the product of the thermal energy per gram carried by the rising bubbles (at constant pressure), 
$C_p \Delta T$ (where $C_p$ is the heat capacity at constant pressure and $\Delta T$ is the
temperature excess between a bubble and its surroundings), and the mass flux, $\frac{1}{2} \rho v_{conv}$
(where $\rho$ and $v_{conv}$ are the local density and convective velocity; the factor 1/2 results from
the assumption that upward and downward flows are identical and, consequently, only half of the
matter moves upward at any time). Accordingly, the convective luminosity can be written as:
\begin{equation}
L_{conv}= 4 \pi r^2 F_{conv} = 2 \pi r^2 \rho v_{conv} C_p \Delta T 
\end{equation}
where $r$ is the location of the convective shell. 

The temperature excess, $\Delta T$, is often expressed in terms of the pressure scale height, $H_P$,
in the form:
\begin{equation}
\Delta T = T \frac{l_m}{H_P} (\nabla - \nabla_{ad})
\end{equation}
$H_P$ is a measure of the characteristic length of the radial variation of the pressure, $P$, hence the explicit mean over a zone, barotropicity,
\begin{equation}
H_P \equiv - \frac{dr}{d ln P} = - P \frac{dr}{dP} = 
    \frac{P}{\rho g}
\end{equation}
for hydrostatic equilibrium conditions
where $g$ is the local gravity. 
$l_m$ is the mixing-length, frequently taken as a multiple of the pressure 
scale height through an adjustable parameter, $\alpha$:
\begin{equation}
l_m = \alpha H_P
\end{equation}

A simple energy balance determines
the convective velocity as a function of the temperature excess and mixing length,
\begin{equation}\label{velof}
v_{conv} = 
 \frac{1}{2^{3/2}} \frac{l_{\rm m}}{r} \left[- \frac{G m}{H_{\rm P}} 
               (\nabla - \nabla_{\rm ad}) 
       \left(\frac{\partial \ln \rho}{\partial \ln T} \right)_{\rm P,\mu} 
       \right]^{1/2} 
\end{equation}
which depends on the choice of the free parameter, $\alpha$.
$\mu$ is the mean molecular weight. 

\begin{figure*}
\centering
\includegraphics[width=6cm, angle=-90]{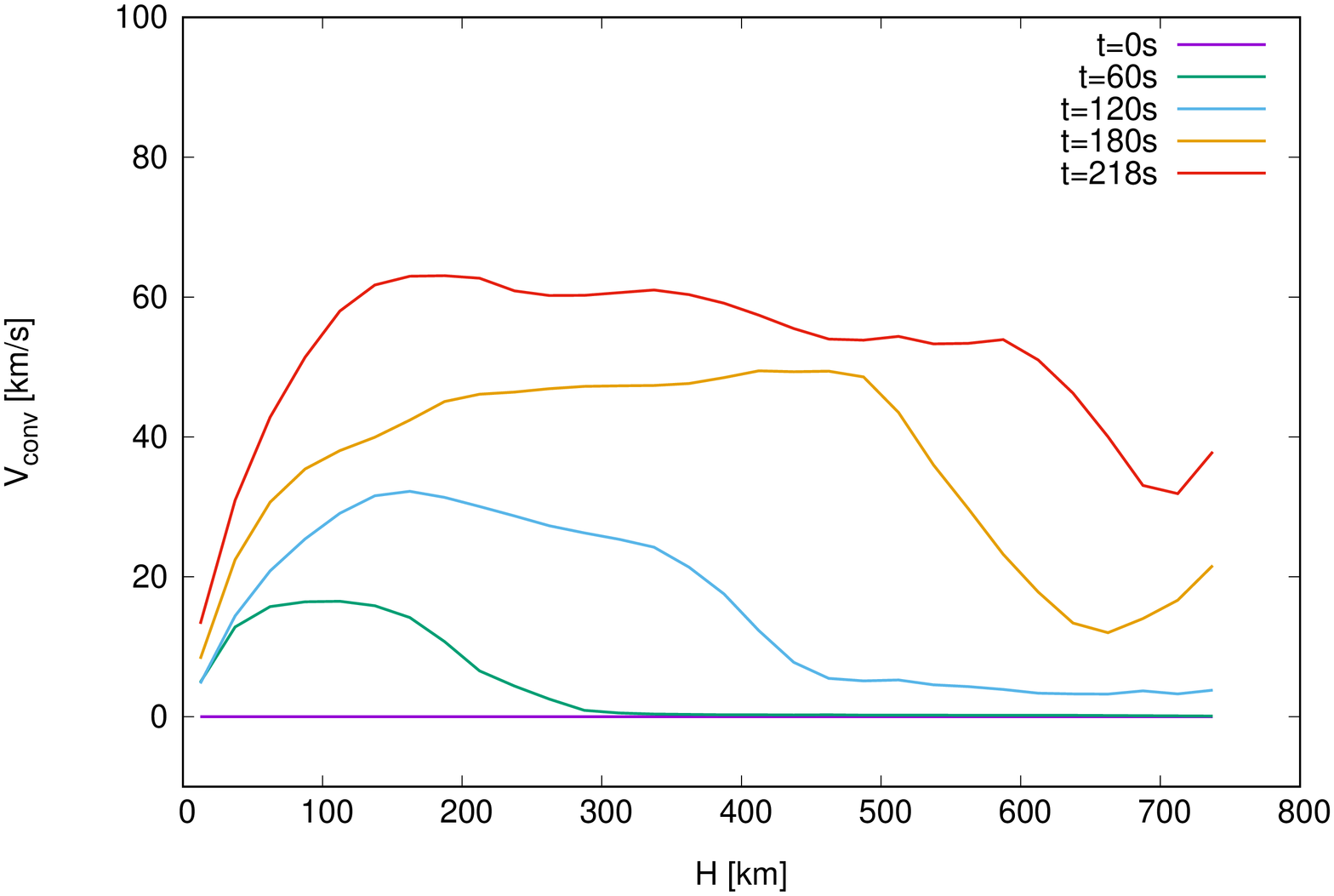}
\includegraphics[width=6cm, angle=-90]{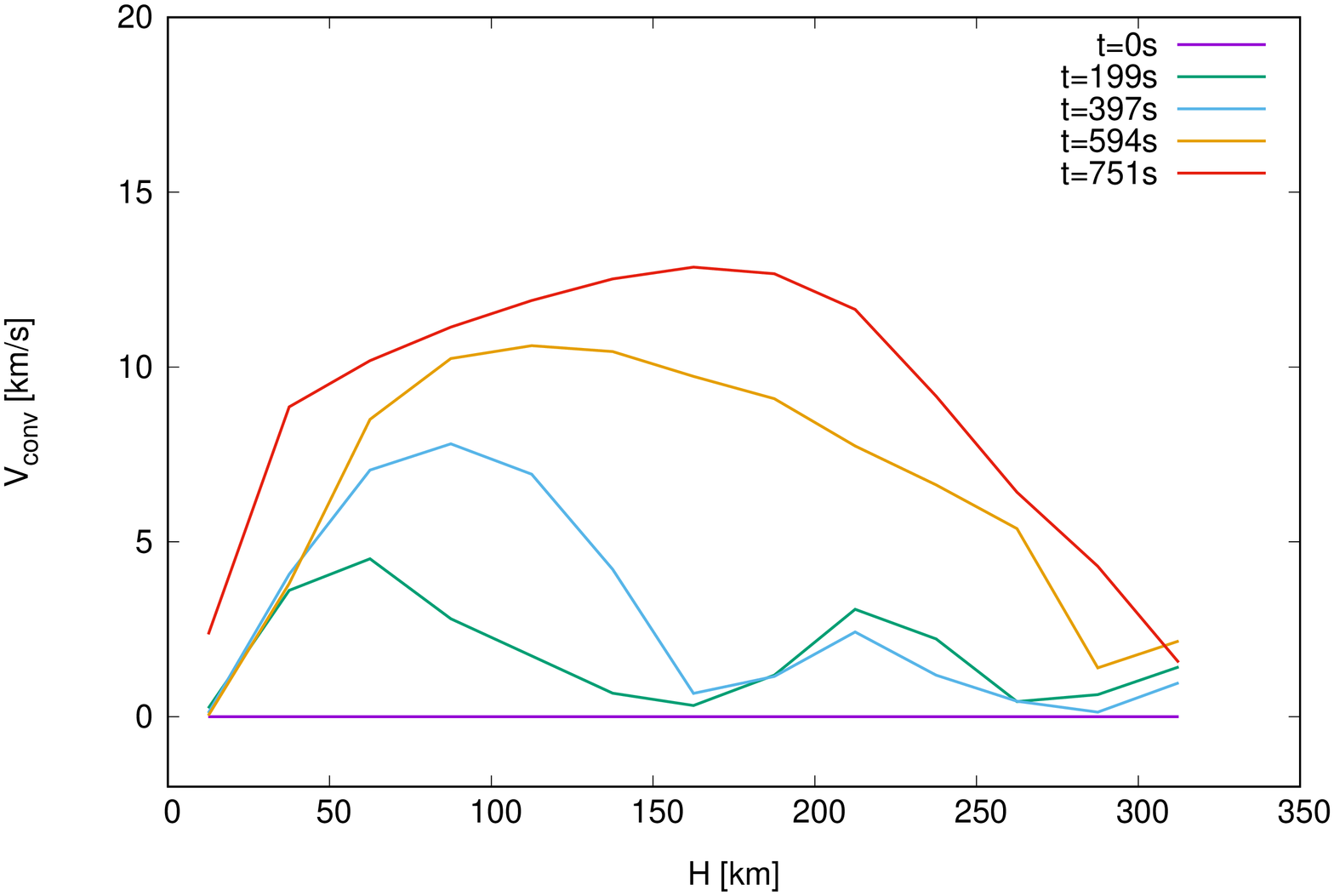}
\caption{Time-dependent, convective velocity profiles across the accreted envelope, extracted
from the 3--D simulations, and used as inputs for 
Models CO1 (left panel) and ONe1 (right panel).}
\label{Fig1}
\end{figure*}

In contrast, in our {\it 321 approach}, the convective velocity profile is directly extracted from the 3--D
simulations as the root mean square of the radial (vertical) component of the velocity vector at
different depths. Therefore, the values of the convective velocity obtained from Eq. \ref{velof} are replaced by 
the values extracted from our 3--D simulations, which are used in turn to determine the convective luminosity through Eq. 1.
It is also worth noting that the 3--D convective velocities, as shown in Fig. \ref{Fig1}, are characterized by a continuous profile,
in sharp contrast with the erratic pattern that results from the application of Schwarzschild criterion in the standard 
mixing-length theory (see, e.g., Fig. \ref{vsound}). In the new {\it 321} models presented in this work, the 3--D convective
velocities are implemented in SHIVA, whether or not Schwarzschild criterion is satisfied (note, however, that convection
almost extends throughout the entire envelope at about $10^8$ K; see Arnett et al. 2015 for a discussion on a 321 approach to 
replace mixing-length theory in stellar evolutionary computations). 
Convective velocity profiles at different times, extracted from the 3--D simulations, are shown in Fig. \ref{Fig1} as a function
of depth, for Models CO1 and ONe1.  
The use of time-dependent convective velocities based on the 3--D simulations directly yields the value of the mixing length, $l_m$,
in a natural way, and independently of any free parameter. 

\begin{figure*}
\centering
\includegraphics[width=6cm, angle=-90]{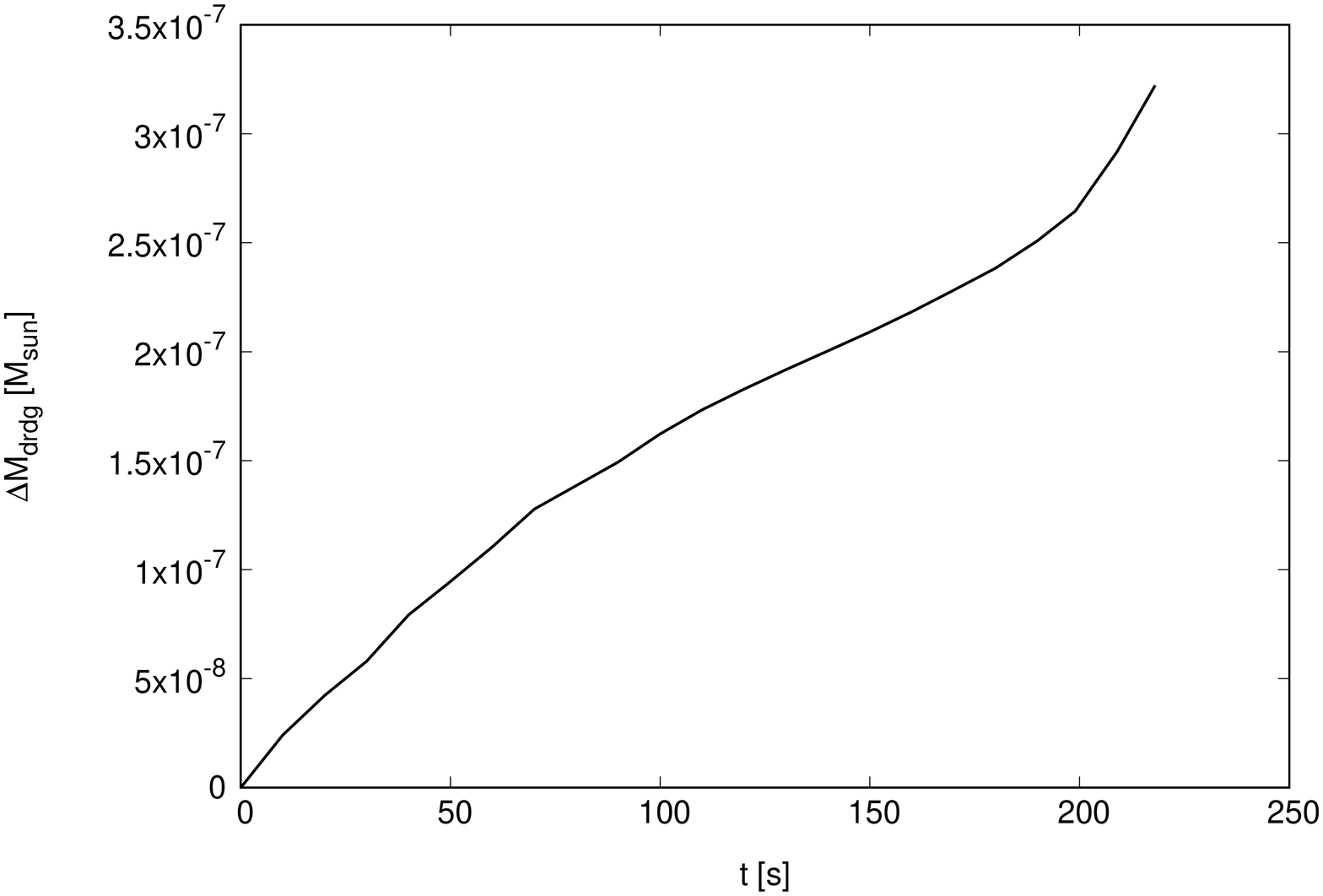}
\includegraphics[width=6cm, angle=-90]{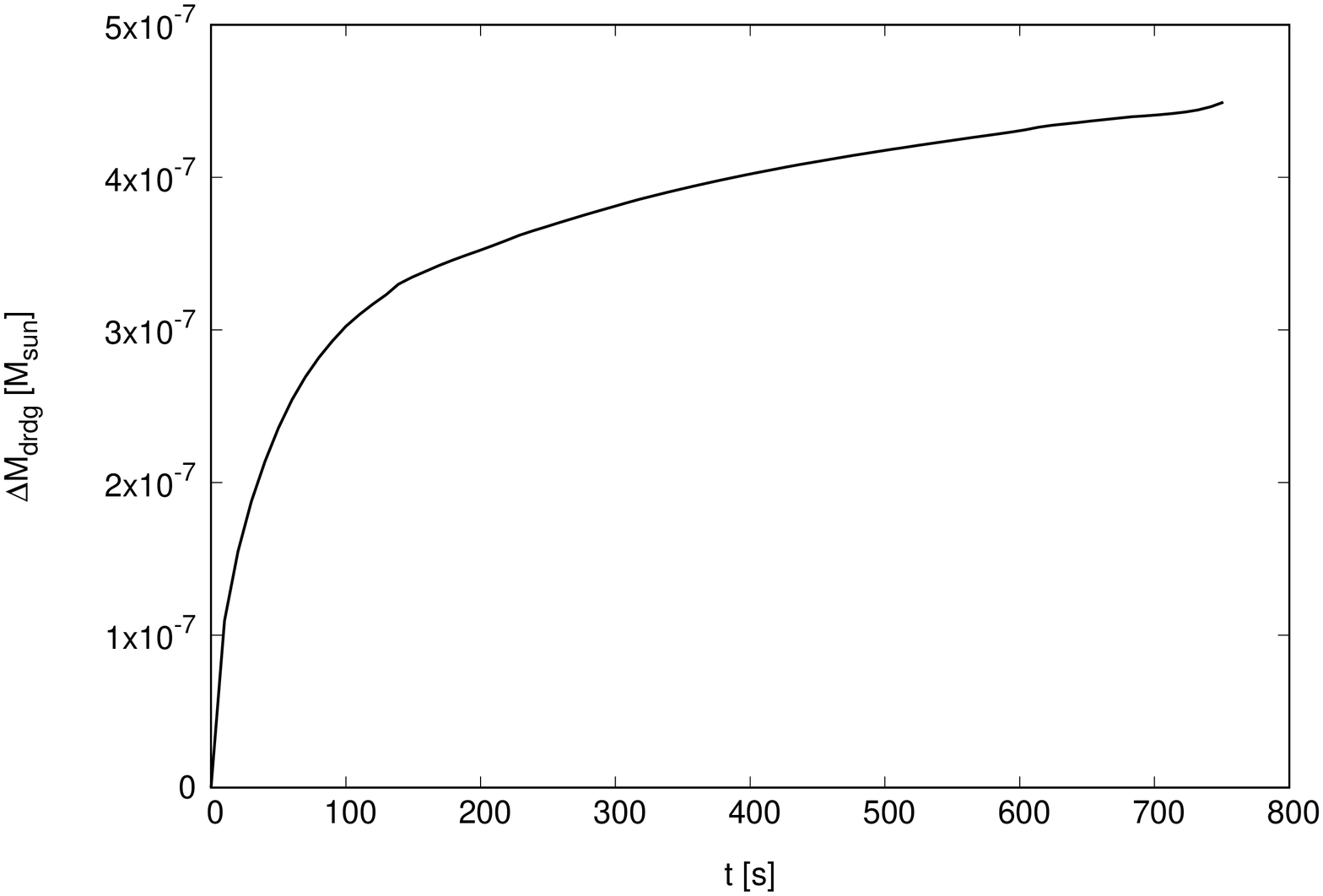}
\caption{Same as Fig. 1, but for the time-dependent mass dredged-up into the accreted envelope.} 
\label{Fig2}
\end{figure*}

The time-dependent amount of mass dredged-up from the outer white dwarf layers extracted from the 3--D simulations (Fig. \ref{Fig2})  has also been included in {\tt SHIVA}.
At each time-step, SHIVA calculates the amount of material dredged-up, on the basis of the 3--D results. This material is mixed with that of 
the innermost envelope shell\footnote{A fully time-dependent stochastic mixing approach, based on the 3--D models, has been developed for post-processing abundance calculations (Leidi 2019a,b).}.
Subsequently, the boundaries of the different numerical shells that characterize the envelope are shifted 
to preserve a constant mass ratio between adjacent shells, and physical variables are interpolated according to the new mass grid (in a similar way in which
mass-accretion is handled).  While responsible for the envelope's metallicity enhancement, dredge-up also
modifies the overall envelope mass and its dynamics. 

\subsection{Models with pre-enrichment}

Different approaches have been adopted to reproduce the
chemical abundance pattern spectroscopically inferred in the  ejecta. 
Traditionally, most 1--D simulations have assumed that the solar-accreted
material is {\it seeded} with material from the outer white dwarf layers, for different percentages
of pre-enrichment (Starrfield et al. 1998, Jos\'e \& Hernanz 1998). While this approach tries to mimic
the mixing at the core-envelope interface, the early enrichment 
of intermediate-mass
elements results in an artificial increase in the envelope's opacity, which in turn affects
the overall amount of mass piled up in the envelope and the dynamics and strength of the subsequent outburst.

To properly assess the implications of our new {\it 123-321} models, we 
computed two additional models with pre-enrichment (hereafter, Models CO2 and ONe2).
In these, the solar-accreted material was seeded with 16\% and 23\% white dwarf material, respectively. 
The adopted percentages correspond to the mean, mass-averaged metallicities in the ejecta
obtained for Models CO1 and ONe1.

\section{Results}

As shown in Table A.1 (see Appendix A), accretion of solar composition material (Z$_{acc} = 0.02$; Model CO1) results in the pile-up of a slightly more massive envelope than
for pre-enriched material (Z$_{acc} = 0.16$; Model CO2). 
As discussed in Jos\'e et al. (2016), the single, most important nuclear reaction during the early stages of a nova outburst is $^{12}$C(p, $\gamma$). 
Its reaction rate is proportional to the product of the mass fractions of the interacting species: for Model CO1, 
this corresponds to 0.711 $\times$ $2.32 \times 10^{-3} = 1.65 \times 10^{-3}$, while for Model CO2,
 0.597 $\times$ $8.20 \times 10^{-2} = 4.90 \times 10^{-2}$. Accordingly, more energy is released per second in Model CO2,
that reaches a thermonuclear runaway earlier than Model CO1. This shortens the overall
duration of the accretion stage in Model CO2, which in turn  reduces the amount of mass accreted in this model. 
Thus, while Model CO1 accretes $8.39 \times 10^{-5}$ M$_\odot$,  Model CO2 accumulates $4.38 \times 10^{-5}$ M$_\odot$ 
(however, since 16\% of the accreted material in Model CO2 corresponds to white dwarf material, the net amount of new material piled up on top of the 
star accounts for only $3.68 \times 10^{-5}$ M$_\odot$). 

As discussed by Shara (1981) and Fujimoto (1982), the strength of a nova outburst is determined by the pressure achieved at the core-envelope interface,
P$_{ce}$, a measure of the overall pressure exerted by the mass overlying the ignition shell, 
\begin{equation}\label{pressio}
P_{ce}= \frac{G M_{wd}}{4 \pi R_{wd}^4} \Delta M_{acc}
\end{equation}
where G is the gravitational constant, M$_{wd}$ and R$_{wd}$ the mass and radius of the white dwarf that hosts the explosion, and
$\Delta M_{acc}$ the mass of the accreted envelope. Mass ejection from the white dwarf surface is achieved for pressures around
P$_{ce} \sim 10^{19}$ - $\sim 10^{20}$ dyn cm$^{-2}$ (Fujimoto 1982, MacDonald 1983). Because of the relationship between
stellar radius and mass, Eq. \ref{pressio} shows that, for a given white dwarf mass, $P_{ce}$ 
depends only on the mass of the accreted envelope. 
Mass accretion onto these CO-rich white dwarfs yields maximum $P_{ce}$ of 
 $1.77 \times 10^{19}$ dyn cm$^{-2}$ (Model CO1) and $9.49 \times 10^{18}$ dyn cm$^{-2}$ (Model CO2).
Thus,  more violent outbursts, characterized by higher peak temperatures,
T$_{max}$, are expected for Model CO1, since it accumulates  about twice as much mass as Model CO2. 
Indeed, Model CO1 achieves T$_{max} = 1.92 \times 10^{8}$ K, while Model CO2 reaches only T$_{max} = 1.72 \times 10^{8}$ K. 

The greater amount of mass accreted in Model CO1 translates into a greater ejected mass, $\Delta M_{eje} = 8.33 
\times 10^{-5}$ M$_\odot$, with a mean kinetic energy of $1.17 \times 10^{45}$ ergs. In contrast, Model CO2 ejects 
$\Delta M_{eje} = 3.54 \times 10^{-5}$ M$_\odot$, with a mean kinetic energy of $5.28 \times 10^{44}$ ergs. 
A mean metallicity of Z$_{eje} = 0.16$ in the ejecta is obtained in Model CO1, mostly as a result of dredge-up from the outermost white dwarf layers. 
In terms of nucleosynthesis, differences in chemical abundances between Models CO1 and CO2 are within a factor of 2 for most species. Notable exceptions
include the light elements $^3$He and $^7$Be ($^7$Li). Their time-evolution is strongly influenced by the longer duration of the
accretion phase in Model CO1. $^{7}$Be, in particular, is a very fragile species. It is synthesized during the early stages of the
runaway by $^3$He($\alpha$, $\gamma$)$^7$Be, and efficiently destroyed by proton-capture reactions, $^7$Be(p, $\gamma$)$^8$B.
However, photodisintegration reactions on $^8$B become important at temperatures above $\geq 10^8$ K, such that a quasi-equilibrium
between the (p, $\gamma$) and the ($\gamma$, p) channels is established, preserving the amount of $^7$Be at this stage
(Hernanz et al. 1996, Jos\'e \& Hernanz 1998).
Therefore, the final amount of $^7$Be in the ejecta, that transforms into $^7$Li by electron captures, critically
depends on the amount available before quasi-equilibrium is reached. This, in turn, depends on the characteristic timescale
of the early runaway: the greater the product X($^1$H) X($^{12}$C), the faster the thermonuclear runaway develops
and the larger the amount of $^7$Li in the ejecta. Accordingly, and as expected, the new {\it 123-321} nova models result
in a net reduction of the $^7$Li produced by novae. Other differences are found in the Mg-Al mass region,
where the abundance of $^{26g}$Al, in particular, is decreased by a factor of $3$ in Model CO1 with respect to the pre-enriched Model CO2.

\begin{figure*}
\centering
\includegraphics[width=6cm, angle=-90]{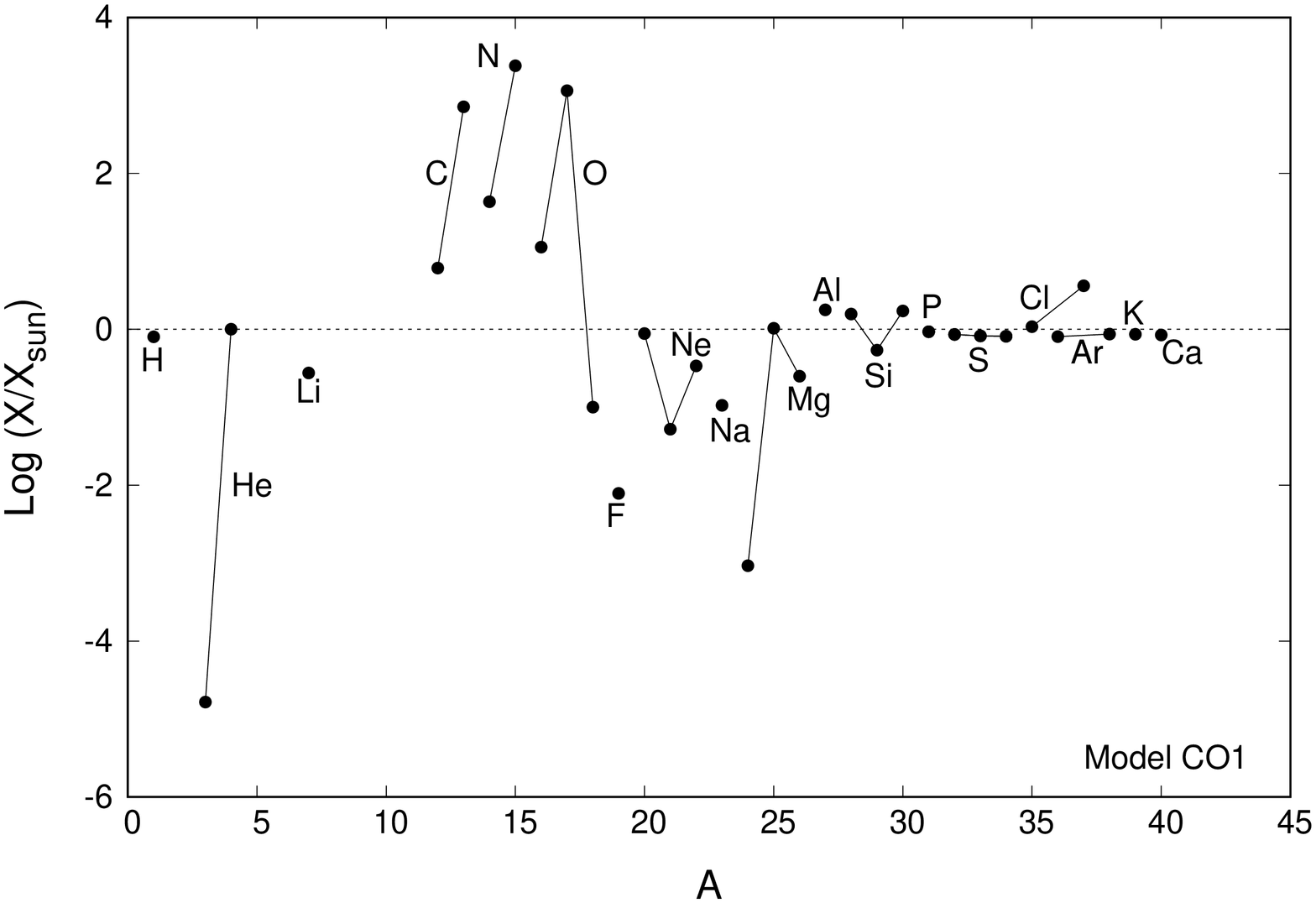}
\includegraphics[width=6cm, angle=-90]{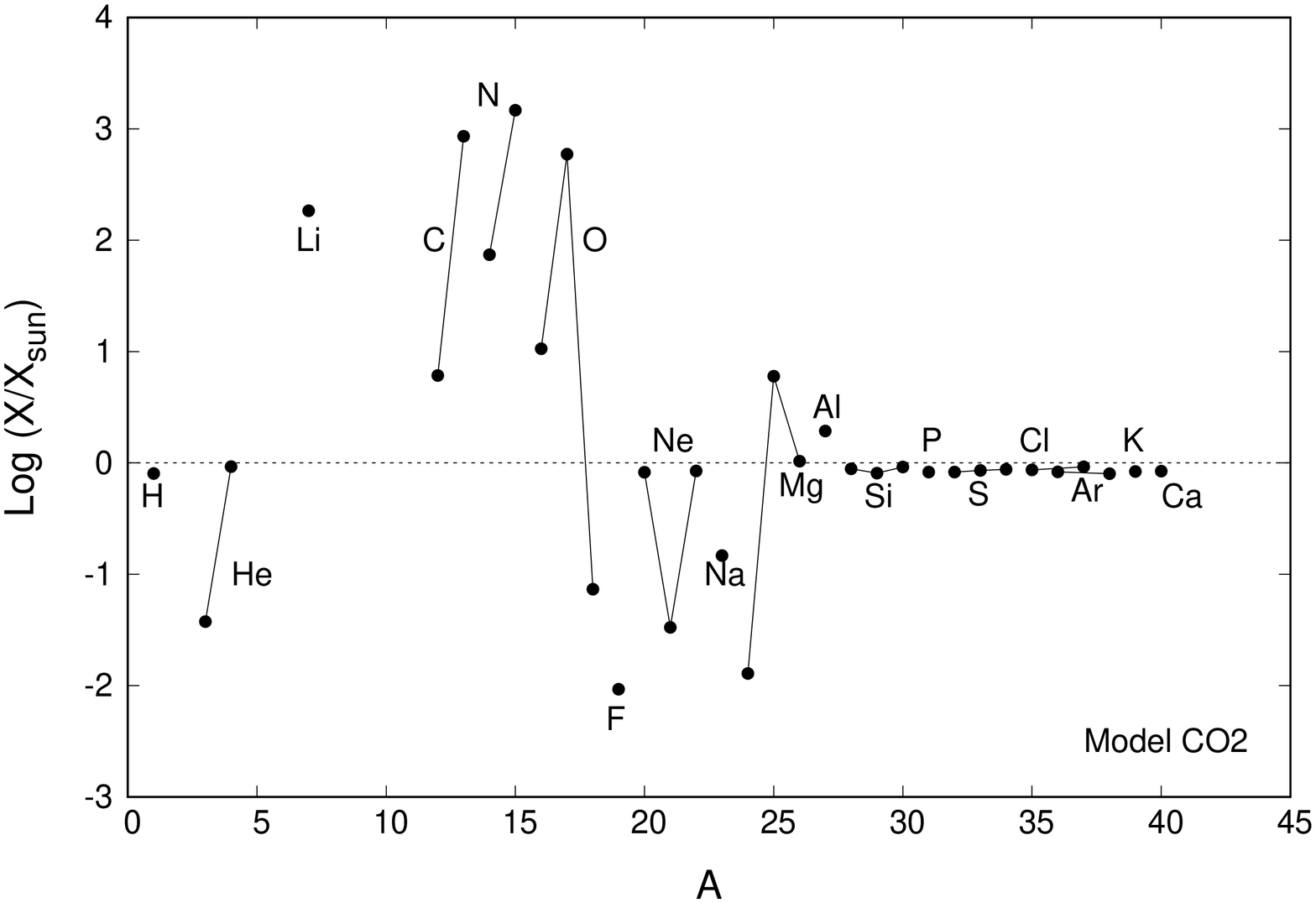}
\caption{Mean overproduction factors, relative to solar (Lodders 2009), in the ejecta of
Models CO1 (left panel) and CO2 (right panel). 
In the plots, unstable nuclei (e.g., $^7$Be)
have been assumed to decay into the corresponding stable species 
(e.g., $^7$Li).}
\label{overp1}
\end{figure*}

\begin{figure*}
\centering
\includegraphics[width=6cm, angle=-90]{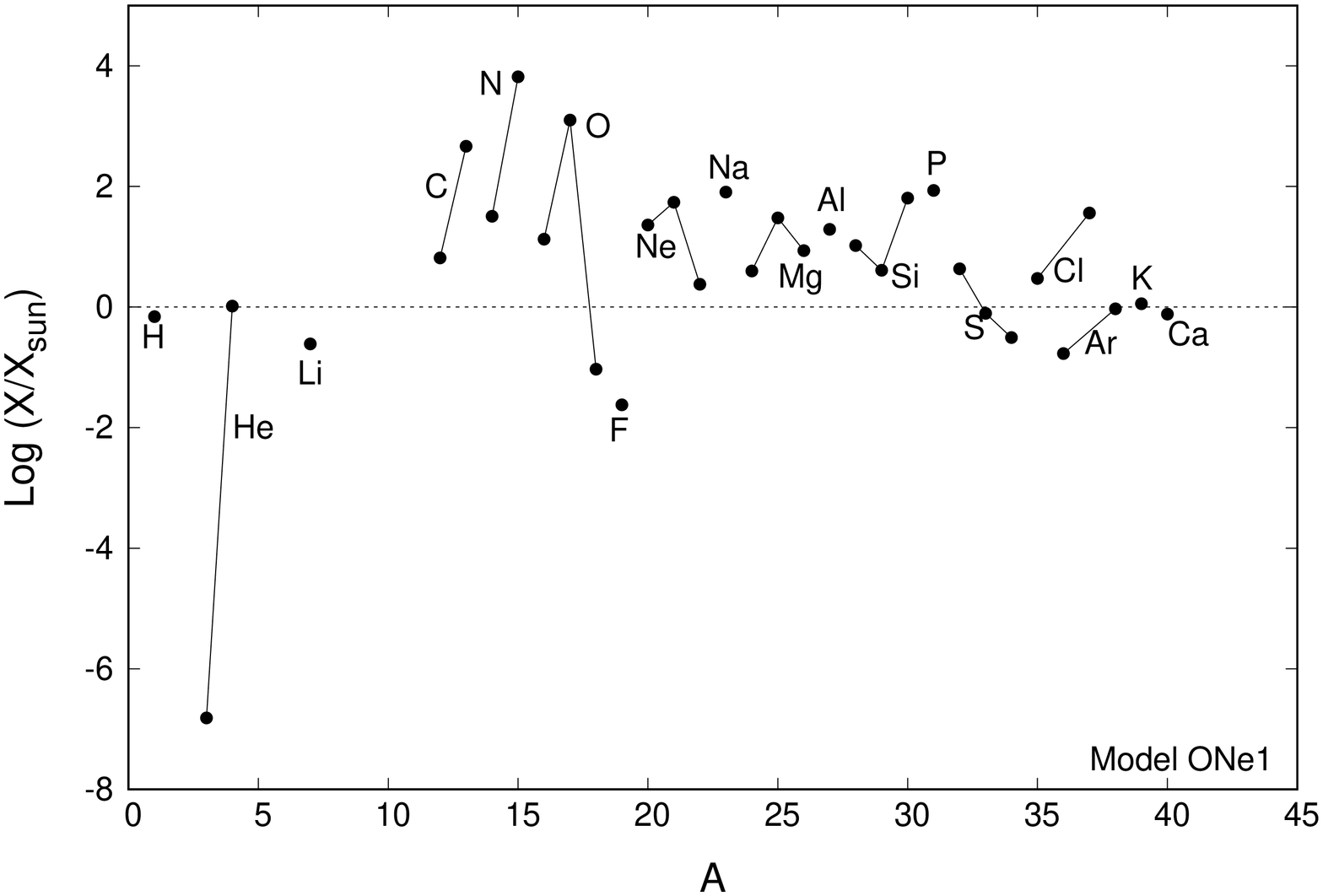}
\includegraphics[width=6cm, angle=-90]{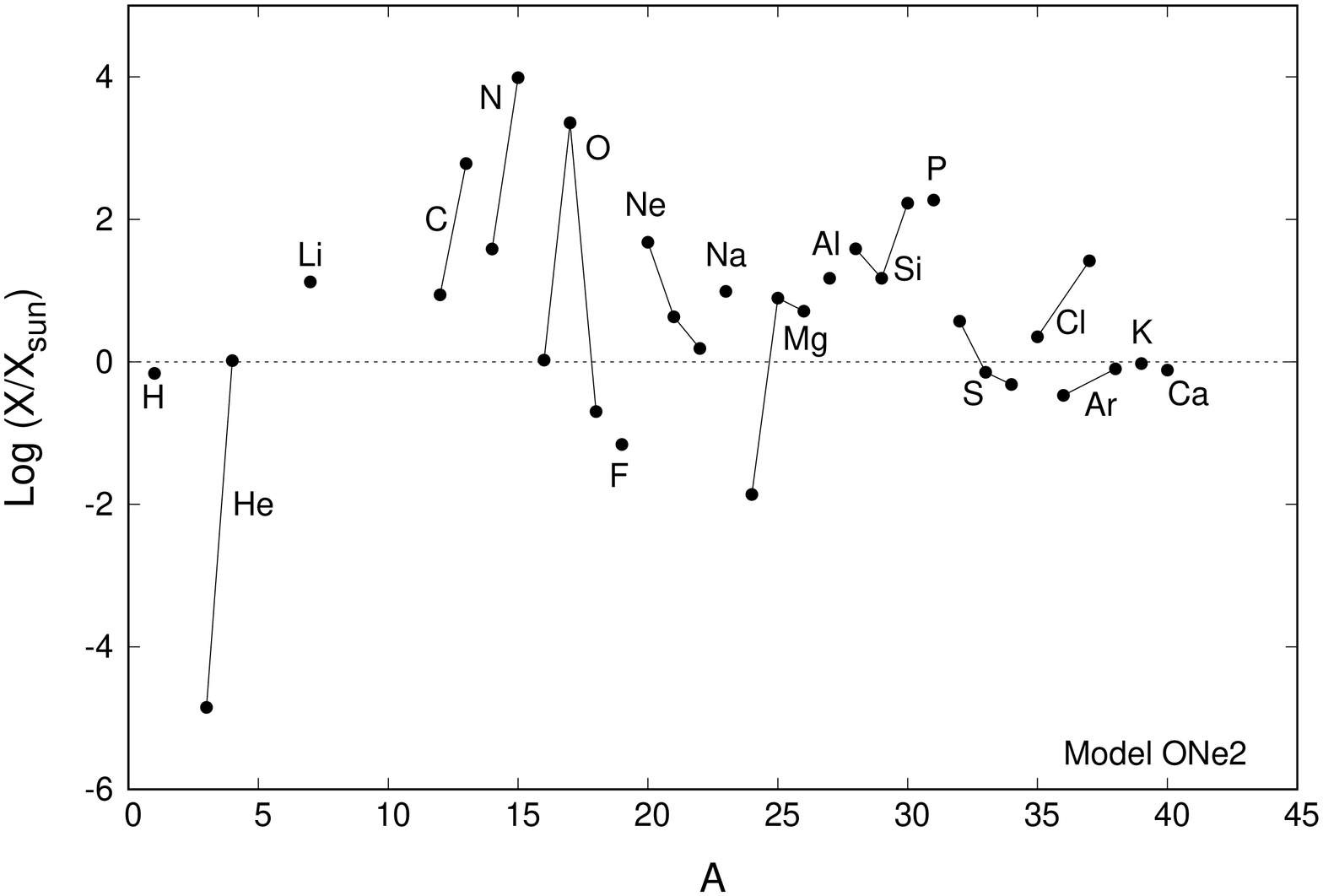}
\caption{Same as Fig. \ref{overp1}, for 
Models ONe1 (left panel) and ONe2 (right panel).}
\label{overp2}
\end{figure*}

Regarding ONe models, 
differences in the overall accreted masses are much smaller than for CO models. 
This is due to similar values of the product X($^1$H) $\times$ X($^{12}$C): 
$1.65 \times 10^{-3}$, as before, for accretion of solar material (Model ONe1), 
and 0.548 $\times$ $3.90 \times 10^{-3}$ = $2.14 \times 10^{-3}$ 
for the pre-enriched Model ONe2. As noted above, the greater the value of the product X($^1$H) $\times$ X($^{12}$C), 
the lower the mass accumulated in the envelope before the thermonuclear runaway sets in.  
Model ONe1 accretes $2.29 \times 10^{-5}$ M$_\odot$, while Model ONe2 accretes $2.12 \times 10^{-5}$ M$_\odot$ 
of pre-enriched material (out of which, $1.63 \times 10^{-5}$ M$_\odot$ was transferred from the stellar companion).
Maximum temperatures and pressures achieved at the core-envelope interface also reflect the similar masses accreted: 
T$_{max} = 2.40 \times 10^8$ K and P$_{ce} = 2.77 \times 10^{19}$ dyn cm$^{-2}$, for Model ONe1;  
T$_{max} = 2.38 \times 10^8$ K and P$_{ce} = 2.59 \times 10^{19}$ dyn cm$^{-2}$, for Model ONe2. 
As found for the CO models, greater ejected masses with higher  kinetic energies were also obtained in the  new ONe model 
with solar accretion and dredge-up: $2.65 \times 10^{-5}$ M$_\odot$ and $1.09 \times 10^{45}$ ergs 
for Model ONe1, while $1.71 \times 10^{-5}$ M$_\odot$ and $0.99 \times 10^{45}$ ergs for Model ONe2.
A mean, mass-averaged metallicity in the ejecta of Z$_{eje} = 0.23$ was obtained in Model ONe1. 
The larger temperatures achieved by the ONe Models extend their nuclear
activity up to Ca. Differences in yields between Models ONe1 and ONe2 reach a factor of $\sim 3$ for many
intermediate-mass elements. Most notably, $^7$Be ($^7$Li) is reduced by a factor $54$ in the new models
with dredge-up (i.e., Model ONe1). 
The mean mass fraction of the $\gamma$-ray emitter $^{22}$Na decreases by a factor of $\sim 2$ in Model ONe1
compared with the value obtained in Model ONe2, while no notable variation is found for $^{26}$Al.
Other nuclear species, such as $^{16}$O, $^{21}$Ne, $^{23}$Na and $^{24,26}$Mg, 
present abundance variations by factors ranging between $8$ and $290$, up and down, when yields 
from Model ONe1 are compared  to values obtained for the pre-enriched Model ONe2 (see Table A.1
and Figs. \ref{overp1} and \ref{overp2}).

\section{Discussion and Conclusions}
\label{sec:conclusions}

This work reports on a new methodology applied to the modeling of nova outbursts, through the use of 1--D and 3--D simulations. It explores, for the first time, the combined effect of mass dredge-up and the inverse energy cascade that characterizes 3--D turbulent convection on the characteristics of the explosion. 

\begin{figure}[bth]
\includegraphics[scale=0.3, angle=-90]{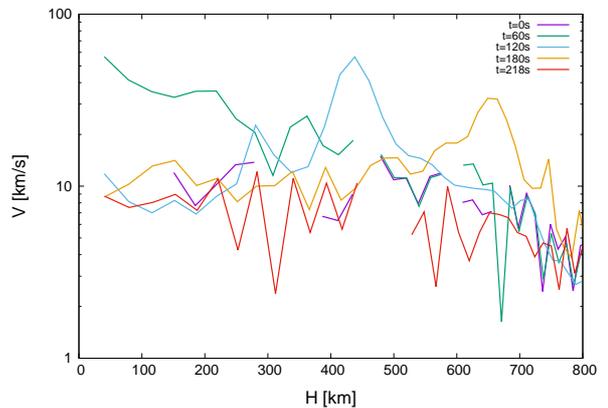}
\caption{Convective velocity 
(derived from  mixing-length theory for $\alpha = 1$) as function
of depth, for the same times displayed in Fig. 1 (left panel; Model CO1). } 
\label{vsound}
\end{figure}

More massive envelopes than those reported from previous models with pre-enrichment have been obtained, providing better agreement with spectroscopically inferred masses.  
The lower estimates systematically predicted by hydrodynamic simulations until now have been regarded as a major drawback of the thermonuclear nova model (see, however, Shore et al. 2016, Mason et al. 2018, for studies of ejected masses based on  reanalyses of filling factors). 
The greater  pressures achieved at the envelope base power, in turn, more violent outbursts, characterized by higher peak temperatures and greater ejected masses, with metallicity enhancements in agreement with observations. 
As shown in Fig. \ref{vsound}, the convective velocities based on our 3--D simulations exhibit  smoother profiles than those based on mixing-length theory (for $\alpha$ = 1). The latter yields more erratic convective patterns, in which convective regions are separated by purely radiative layers. Moreover, maximum convective velocities obtained in our 3--D simulations exceed the estimates based on mixing-length theory when convective transport almost extends through the entire envelope (i.e., at 120 s and 218 s). Such convective velocites are distinctly subsonic,  as can  be shown by comparison with the values of the local speed of sound, $c_s$,  calculated as:
\begin{equation}\label{csound}
c_{s} = 
 \sqrt{\left(\frac{\partial P}{\partial \rho} \right)_{\rm S}} =  
 \sqrt{ \left(\frac{\partial P}{\partial \rho} \right)_{\rm T} +         
 \frac{T}{\rho^2} 
 \frac{ \left(\frac{\partial P}{\partial T} \right)^2_{\rm \rho}}{\left(\frac{\partial U}{\partial T} \right)_{\rm \rho}}          }   
\end{equation}
where $S$ is the entropy and $U$ is the internal energy per unit mass.

Due to the Eulerian nature of the {\tt FLASH} code, our 3--D simulations must be stopped when the convective front hits the upper boundary of the computational domain. This limits the extent of the multidimensional simulations, that cannot proceed
through most of the expansion and ejection stages. 
Therefore, prescriptions for the time-dependent amount of mass dredged-up and convective velocity are not available much
beyond the peak of the explosions. To overcome this limitation, in the {\it 321} simulations reported in this work, mass dredge-up is extrapolated from the last
data points available from the 3--D simulations. 
With regard to the time-dependent convective
velocity, no extrapolation is made. Instead, {\tt SHIVA} switches to a  time-dependent mixing length prescription at longer times, to properly account for the progressive retreat of convection from the outermost envelope layers. 
 It is worth noting that the general trends reported in this paper 
(i.e., larger accreted and ejected masses, favoring a better agreement with the values inferred spectroscopically; more violent outbursts, characterized
by larger peak temperatures and kinetic energies; a reduction of the overall amount of $^7$Be ($^7$Li) in the ejecta)
are not affected by the extrapolation of the time-dependent mass dredged-up 
into the envelope adopted during the last stages of the outburst. However, some specific details, such as the exact amount of mass
ejected in the outburst, or the final mass fractions of some nuclear species in the ejecta (particularly those dredged-up from the outermost white dwarf layers) would
be affected, to some extent, by the way the mass dredged-up proceeds during the last stages of the explosion.
It is also important to stress that the last stages of the outburst are characteried by decaying turbulence,
  and therefore mixing should persist after the explosion.
Efforts aimed at extending the computational domains used in our 3--D simulations are currently underway.
These should allow {\tt FLASH} to proceed further, until the envelope becomes physically detached from
the rest of the star, and dredge-up of white dwarf material and convection are naturally halted. 

Finally, the simulations reported in this work suggest that the white dwarf mass decreases after a nova outburst,  
whenever metallicity enhancements in the ejecta exceed a threshold value around $Z_{eje} \sim 0.2$. This has implications for the long-debated role of classical novae as possible type Ia supernova progenitors, since the metallicities inferred from the nova ejecta frequently exceed that threshold.

{\it Acknowledgements.} 
The authors would like to thank Margarita Hernanz, Christian Iliadis, Giovanni Leidi, and Sumner Starrfield
for fruitful discussions. 
 This work has been partially
supported by the Spanish MINECO grant AYA2017--86274--P,
 by the E.U. FEDER funds, and by the AGAUR/Generalitat de Catalunya grant 
SGR-661/2017. This article benefited from discussions within the ``ChETEC'' COST Action (CA16117).

\begin{appendix}
\section{Nucleosynthesis}

\begin{table*}[htb]
\caption{Nova models computed in this work. 
} 
    \label{Series1}
    \centering   
    \begin{tabular}{l c c c c}
    \hline\hline\
% 
%  Model CO1: *.test1 up to T8_ce = 1 (/321_100CO) + *.3dfullCO1b (/321_100COfull3D) 
%  Model CO2: *.test16 (/321_100CO)
%  Model ONe1: *.test1 up to T8_ce = 1 (/321_125ONe) + *.div1 (/321_125ONefull3D)
%  Model ONe2: *.test27 (/321_125ONe)
%  Files: 3dfullCO1b.mean (CO1), test16.mean (CO2), div1.mean (ONe1), test27.mean (ONe2)
    Model                                 & CO1                 & CO2                 & ONe1                 & ONe2 \\                  
    Type                                  & 123-321               & 1--D with pre-enrichment & 123-321           & 1--D with pre-enrichment \\                  
    M$_{\rm WD}$($M_\odot$)               & 1.0                 & 1.0                 & 1.25                 & 1.25 \\ 
    Z$_{acc}$                             & Solar               & 16\% CO, 84\% Solar & Solar                & 23\% ONe, 77\% Solar \\
    \hline
    $\Delta$M$_{acc}$($10^{-5}$ M$_\odot$)& 8.39                & 3.68$^a$            & 2.29                 & 1.63$^a$    \\
    T$_{max}$($10^8$ K)                   & 1.92                & 1.72                & 2.40                 & 2.38        \\
    K($10^{45}$ ergs)                     & 1.17                & 0.528               & 1.09                 & 0.99        \\
    v(km s$^{-1}$)                        & 1190                & 1240                & 2030                 & 2410        \\
    $\Delta$M$_{eje}$($10^{-5}$ M$_\odot$)& 8.33                & 3.54                & 2.65                 & 1.71        \\
    Z$_{eje}$                             & 0.16                & 0.18                & 0.23                 & 0.23        \\
    X($^1$H)                              & 5.7(-1)             & 5.7(-1)             & 4.9(-1)              & 4.9(-1)     \\
    X($^3$He)                             & 1.4(-9)             & 3.2(-6)             & -                    & 1.2(-9)     \\
    X($^4$He)                             & 2.7(-1)             & 2.5(-1)             & 2.8(-1)              & 2.8(-1)     \\
    X($^7$Li)                             & -                   & 1.3(-9)             & -                    & -           \\
    X($^7$Be)                             & 2.7(-9)             & 1.8(-6)             & 2.4(-9)              & 1.3(-7)     \\
    X($^{12}$C)                           & 1.4(-2)             & 1.4(-2)             & 1.5(-2)              & 2.0(-2)     \\
    X($^{13}$C)                           & 2.0(-2)             & 2.4(-2)             & 1.3(-2)              & 1.7(-2)     \\
    X($^{14}$N)                           & 3.5(-2)             & 6.0(-2)             & 2.6(-2)              & 3.1(-2)     \\
    X($^{15}$N)                           & 7.7(-3)             & 4.7(-3)             & 2.1(-2)              & 3.1(-2)     \\
    X($^{16}$O)                           & 7.7(-2)             & 7.2(-2)             & 9.1(-2)              & 7.2(-3)     \\
    X($^{17}$O)                           & 3.1(-3)             & 1.6(-3)             & 3.4(-3)              & 6.1(-3)     \\
    X($^{18}$O)$^b$                       & 3.4(-7)             & 2.5(-7)             & 3.8(-7)              & 8.9(-7)     \\
    X($^{18}$F)$^b$                       & 1.2(-6)             & 8.2(-7)             & 1.0(-6)              & 2.1(-6)     \\
    X($^{19}$F)                           & 3.3(-9)             & 3.9(-9)             & 1.0(-8)              & 2.9(-8)     \\
    X($^{20}$Ne)                          & 1.5(-3)             & 1.4(-3)             & 3.9(-2)              & 8.1(-2)     \\
    X($^{21}$Ne)                          & 2.2(-7)             & 1.4(-7)             & 2.3(-4)              & 1.8(-5)     \\
    X($^{22}$Ne)                          & 4.4(-5)             & 1.1(-4)             & 2.8(-4)              & 1.3(-4)     \\
    X($^{22}$Na)                          & 8.5(-7)             & 7.3(-7)             & 3.2(-5)              & 6.5(-5)     \\
    X($^{23}$Na)                          & 3.8(-6)             & 5.3(-6)             & 2.9(-3)              & 3.5(-4)     \\
    X($^{24}$Mg)                          & 4.9(-7)             & 6.8(-6)             & 2.1(-3)              & 7.3(-6)     \\
    X($^{25}$Mg)                          & 7.2(-5)             & 4.2(-4)             & 2.1(-3)              & 5.5(-4)     \\
    X($^{26}$Mg)                          & 5.6(-6)             & 4.2(-5)             & 5.1(-4)              & 2.7(-5)     \\
    X($^{26g}$Al)                         & 1.4(-5)             & 4.1(-5)             & 1.8(-4)              & 1.4(-4)     \\
    X($^{27}$Al)                          & 1.1(-4)             & 1.2(-4)             & 1.2(-3)              & 9.2(-4)     \\
    X($^{28}$Si)                          & 1.1(-3)             & 6.2(-4)             & 7.3(-3)              & 2.7(-2)     \\
    X($^{29}$Si)                          & 2.0(-5)             & 3.0(-5)             & 1.5(-4)              & 5.5(-4)     \\
    X($^{30}$Si)                          & 4.3(-5)             & 2.3(-5)             & 1.6(-3)              & 4.2(-3)     \\
    X($^{31}$P)                           & 6.5(-6)             & 5.8(-6)             & 6.0(-4)              & 1.3(-3)     \\
    X($^{32}$S)                           & 3.0(-4)             & 2.9(-4)             & 1.5(-3)              & 1.3(-3)     \\
    X($^{33}$S)                           & 2.3(-6)             & 2.4(-6)             & 2.2(-6)              & 2.0(-6)     \\
    X($^{34}$S)                           & 1.3(-5)             & 1.4(-5)             & 5.0(-6)              & 7.7(-6)     \\
    X($^{35}$Cl)                          & 4.0(-6)             & 3.2(-6)             & 1.1(-5)              & 8.3(-6)     \\
    X($^{36}$Ar)                          & 6.2(-5)             & 6.4(-5)             & 1.3(-5)              & 2.6(-5)     \\
    X($^{37}$Cl)                          & 4.7(-6)             & 1.2(-6)             & 4.7(-5)              & 3.4(-5)     \\
    X($^{38}$Ar)                          & 1.3(-5)             & 1.2(-5)             & 1.4(-5)              & 1.2(-5)     \\
    X($^{39}$K)                           & 3.2(-6)             & 3.1(-6)             & 4.2(-6)              & 3.5(-6)     \\
    X($^{40}$Ca)                          & 5.4(-5)             & 5.4(-5)             & 4.9(-5)              & 4.9(-5)     \\
    \hline
    \end{tabular}
\vspace{0.1 cm}
\begin{list}{}{} 
\item[$^{\mathrm{a}}$] Fraction of the accreted mass that corresponds to solar composition (84\% and 77\%, respectively), excluding pre-enrichment with white dwarf material. 
\item[$^{\mathrm{b}}$] Mass fractions correspond to t = 1 hr after T$_{max}$.
\end{list}
    \end{table*}

\end{appendix}
\end{document}